\documentclass[showpacs, preprintnumbers, nofootinbib, aps, prd,
superscriptaddress,10pt, showkeys, notitlepage,twocolumn]{revtex4-1}

\usepackage{graphicx,amssymb,amsmath,amsthm,amsfonts,epsfig}

\usepackage[linktocpage]{hyperref}
\usepackage[usenames,dvipsnames]{color}
\usepackage{epstopdf}
\usepackage{aas_macros}
\usepackage{pifont}
\definecolor{darkred}{rgb}{0.5,0,0}
\definecolor{darkgreen}{rgb}{0,0.5,0}
\definecolor{darkblue}{rgb}{0,0,0.5}
\definecolor{prussian}{rgb}{0.0, 0.19, 0.33}
\definecolor{richelectricblue}{rgb}{0.03, 0.57, 0.82}
\definecolor{teal}{rgb}{0.0, 0.5, 0.5}
\definecolor{mediumseagreen}{rgb}{0.24, 0.7, 0.44}
\definecolor{lust}{rgb}{0.9, 0.13, 0.13}
\definecolor{ballblue}{rgb}{0.13, 0.67, 0.8}
\definecolor{darkcyan}{rgb}{0.0, 0.55, 0.55}
\definecolor{mountainmeadow}{rgb}{0.19, 0.73, 0.56}
\definecolor{palecarmine}{rgb}{0.69, 0.25, 0.21}
\definecolor{richcarmine}{rgb}{0.84, 0.0, 0.25}
\definecolor{tangelo}{rgb}{0.98, 0.3, 0.0}
\definecolor{venetian}{rgb}{0.784,0.031,0.082}
\definecolor{bdfrance}{rgb}{0.192,0.549,0.906}

\hypersetup{colorlinks=true,
            citecolor=venetian,
            linkcolor=NavyBlue,
            urlcolor=palecarmine}

\usepackage[utf8]{inputenc}
\usepackage{amsmath,amssymb}
\usepackage{tensor}
\usepackage{mathtools}
\usepackage{amsbsy}
\usepackage{bm}
\usepackage{float}
\newcommand{\be}{\begin{equation}}
\newcommand{\ee}{\end{equation}}
\newcommand{\bear}{\begin{eqnarray}}
\newcommand{\eear}{\end{eqnarray}}

\newcommand{\p}{\prime}
\newcommand{\pp}{\prime\prime}
\newcommand{\nn}{\nonumber}

\newcommand{\cB}{{\cal B}}
\newcommand{\cE}{{\cal E}}

\newcommand{\cD}{{\cal D}}

\newcommand{\cC}{{\cal C}}
\newcommand{\cA}{{\cal A}}

\newcommand{\cW}{{\cal W}}
\newcommand{\cF}{{\cal F}}
\newcommand{\cG}{{\cal G}}
\newcommand{\cQ}{{\cal Q}}
\newcommand{\cH}{{\cal H}}

\newcommand{\rM}{{\rm m}}

\newcommand{\vt}{\tilde{V}}

\newcommand{\ddSm}{(S_{,xx})_{\rm m}}

\newcommand{\order}[1]{{\cal O}(\ell^{#1})}

\begin{document}

\title{Eikonal quasinormal modes of black holes beyond general relativity II: \\ generalised scalar-tensor perturbations}

\author{Hector O. Silva}
\email{hosilva@illinois.edu}
\affiliation{eXtreme Gravity Institute, Department of Physics, Montana State University, Bozeman, Montana 59717, USA}
\affiliation{Department of Physics, University of Illinois at Urbana-Champaign, Urbana, Illinois 61801, USA}

\author{Kostas Glampedakis}
\email{kostas@um.es}
\affiliation{Departamento de F\'isica, Universidad de Murcia, Murcia, E-30100, Spain}
\affiliation{Theoretical Astrophysics, University of T\"ubingen, Auf der Morgenstelle 10, T\"ubingen, D-72076, Germany}

\begin{abstract}

Black hole `spectroscopy', i.e. the identification of quasinormal mode
frequencies via gravitational wave observations, is a powerful technique for
testing the general relativistic nature of black holes. In theories of gravity
beyond general relativity perturbed black holes are typically described by a
set of coupled wave equations for the tensorial field and the extra
scalar/vector degrees of freedom, thus leading to a theory-specific quasinormal
mode spectrum. In this paper we use the eikonal/geometric optics approximation
to obtain analytic formulae for the frequency and damping rate of the
fundamental quasinormal mode of a generalised, theory-agnostic system of
equations describing coupled scalar-tensor perturbations of spherically
symmetric black holes. Representing an extension of our recent work, the
present model includes a massive scalar field, couplings through the field
derivatives and first-order frame-dragging rotational corrections. Moving away
from spherical symmetry, we consider the simple model of the scalar wave
equation in a general stationary-axisymmetric spacetime and use the eikonal
approximation to compute the quasinormal modes associated with equatorial and
nonequatorial photon rings.

\end{abstract}

\maketitle

%%%%%%%%%%%%%%%%%%%%%%%%%%%%%%%%%%%%%%%%%%%%%%%%%%%%%%%%%%%%%%%%%%%%%%%%%

\section{Introduction}
\label{sec:intro}

% ------------------------
% Gravitational wave astro
% ------------------------
The first direct detections of gravitational waves (GWs) from coalescing binary
black holes by the LIGO and Virgo collaboration have marked the beginning of the GW
astronomy era~\cite{Abbott:2016nmj,Abbott:2016blz,Abbott:2017vtc}.
This new form of astronomy has allowed us to test general relativity (GR), and
to constrain modifications to it, in a regime where the theory's full
nonlinearity is manifest for the first time, see
e.g.~\cite{Yunes:2016jcc,LIGOScientific:2019fpa,Nair:2019iur,Tahura:2019dgr}.

% --------------
% Ringdown tests
% --------------
One of the main goals of present and future GW observatories is to detect the
`dying GW notes' emitted by the merger remnant as it relaxes to its final
state.
This `ringdown' is characterized by a set of complex-valued frequencies known
as quasinormal modes (QNMs), which depend only on the properties of the final
black hole, that is, its mass and spin.
The detection of two or more QNMs would allow us to do black hole
`spectroscopy'~\cite{Detweiler:1980gk}, that is, to infer the properties of the remnant
black hole by its QNM spectra, analogously to how atomic elements can be
identified by their emission spectra~\cite{Dreyer:2003bv,Berti:2005ys,Berti:2007zu,Berti:2016lat,Yang:2017zxs,Yang:2017xlf,Isi:2019aib}.

From a fundamental physics standpoint, the most appealing aspect of black hole
spectroscopy is the prospect of testing the `Kerr hypothesis', which states
that all astrophysical BHs are described by the Kerr metric~\cite{Cardoso:2016ryw}.
A violation of this hypothesis would hint towards new physics beyond GR~\cite{Berti:2018vdi}.
This has motivated considerable effort in modeling the QNM spectrum of black
holes outside the Kerr paradigm, ranging from model-independent
parametrisations~\cite{Glampedakis:2017,Tattersall_etal2018,Cardoso_etal2019,McManus:2019ulj,Maselli:2019mjd,Volkel:2019muj}
to calculations done in specific beyond-GR theories, e.g.~\cite{Ferrari:2000ep,Molina:2010fb,Kobayashi:2012kh,Kobayashi:2014wsa,
Blazquez-Salcedo:2016enn,Brito:2018hjh,Tattersall_2018a,Tattersall:2019pvx}.

In this paper we continue our programme started in~\cite{Glampedakis:2019dqh}
(hereafter `Paper I'), where we used the eikonal/geometric optics limit to
calculate analytically the quasinormal modes associated with coupled
tensor-scalar systems of wave equations which arise naturally in theories
beyond GR.
Our goal is to extend Paper I in two important directions: (i) by
studying a more general system of equations and (ii) by including the
leading-order slow rotation corrections.
We also consider stationary-axisymmetric spacetimes, where for simplicity,
we consider massless scalar field perturbations.
We use the eikonal limit to study the the QNMs of massless scalar perturbations
associated with equatorial and nonequatorial photon rings which arise in certain
non-Kerr geometries.

The rest of this paper is organized as follows.
In Sec.~\ref{sec:coupledEqs} we introduce the coupled system of perturbation
equations studied in this paper.
In Sec.~\ref{sec:leading_order} we consider the leading-order eikonal formulae
that arises from this system and in Sec.~\ref{sec:Veff} we introduce an
effective potential associated with these perturbations.
The QNM calculation is completed in Sec.~\ref{sec:subleading_order}, where we
consider subleading order eikonal QNM corrections.
These sections constitute the main calculation of this paper and the main
results are summarized in Sec.~\ref{sec:summary}.
In the two following sections we relax some of the assumptions made in the
previous sections. First, in Sec.~\ref{sec:exotic} we explore the possibility
of noncanonical eikonal QNMs. Next,
in Sec.~\ref{sec:nonspherical}, we study the scalar eikonal QNMs in generic
axisymmetric and stationary backgrounds.
% %
Our conclusions and possible directions for further work are found in
Sec.~\ref{sec:conclusions}.

Throughout this paper we use geometric units $G = c = 1$. Primes stand for
radial derivatives $d/dr$.  For any function $f(r)$ we use the abbreviation $f_{z}
\equiv f(r_z)$.
When enclosed within brackets, numeric indices label the eikonal order of a
given function, otherwise they represent slow-rotation expansions.

%%%%%%%%%%%%%%%%%%%%%%%%%%%%%%%%%%%%%%%%%%%

\section{The coupled wave equations}
\label{sec:coupledEqs}

We consider perturbed non-GR black holes described by a general system of wave
equations with two coupled scalar-tensor field degrees of freedom $\{\Theta, \psi\}$.
Our equations are supposed to be theory-agnostic but they do include,
for instance, the perturbation equations of Schwarzschild black holes within
the framework of the generalised class of scalar-tensor theories discussed
in~\cite{Tattersall_etal2018}.
Although we assume spherically symmetric black holes, we can also allow for the
possibility of first-order rotational corrections. After the separation of the
angular part and assuming a $\sim e^{-i\omega t}$ time dependence, the system
of coupled equations takes the form,
\begin{subequations}
\begin{align}
&\frac{d^2 \psi}{dx^2} + \left [\, \omega^2 -2m\omega \Omega(r) - V_\psi(r) \, \right ] \psi = \beta_\psi (r) \Theta,
\label{waveT}
\\
& \frac{d^2 \Theta}{dx^2} + g(r)  \frac{d\Theta}{dx} + \left [\, \omega^2 -2m\omega \Omega(r) - V_\Theta (r) \, \right ] \Theta =
\nn \\
& = \beta_\Theta (r) \psi + b_1 (r) \frac{d\psi}{dx} + b_2 (r) \frac{d^2 \psi}{dx^2}.
\label{waveS}
\end{align}
\label{sysWaves}
\end{subequations}
We have denoted with $x=x(r)$ the tortoise coordinate that eliminates the first-order
radial derivative in the equation for the tensorial field $\psi$.
The potential $V_\psi$ in that equation can be assumed to be identical to the
Schwarzschild's spacetime Regge-Wheeler or Zerilli potential while the
potential for the scalar field $\Theta$ (to which we assign a mass
$\mathfrak{m}$ with associated inverse Compton wavelength
$\mu \equiv \mathfrak{m} / \hbar$) is allowed to deviate from GR:
\begin{align}
V_\psi &= \{ V_{\rm RW}, V_{\rm Z} \},
%f(r) \left [\, \frac{\ell(\ell+1)}{r^2} - \frac{6M}{r^3} \, \right ],
\\
V_\Theta &= f(r) \left [\, \frac{\ell(\ell+1)}{r^2}\alpha(r) + \frac{2M}{r^3} \zeta (r) + \mu^2 \, \right ].
\end{align}
We further assume that the functions $\{f, g,\alpha,\zeta \}$ carry no $\ell$ or $\omega$ dependence.
In contrast, the coupling functions $\{ \beta_\psi, \beta_\Theta, b_1, b_2\}$ are \emph{a priori} expected to depend on both of these parameters.
For later convenience we define,
\be
\beta_{\psi\Theta} = \beta_\psi \beta_\Theta, \quad \gamma_{\psi\Theta} =  \beta_\psi b_1, \quad
\delta_{\psi\Theta} =  \beta_\psi b_2.
\label{defcouplings}
\ee
The rotational terms $-2m\omega \Omega$ appearing in the above equations have been added `by hand' and correspond
to the leading-order spin term of the scalar wave equation in GR, assuming  a stationary-axisymmetric spacetime with a
$g_{t\varphi} = - \Omega r^2 \sin^2\theta $ metric component (in GR the `frame-dragging' angular frequency turns out to
be $\Omega = 2 J/r^3$ where $J$ is the black hole's angular momentum~\cite{MTW1973}). As a disclaimer, it should be pointed
out that the possible presence of rotational corrections in the coupling terms has not been taken into account here. We should also
note that the leading-order spin term in the gravitational equation is expected to be significantly more complicated  than the one assumed
here; this can be demonstrated by taking the slow-rotation limit of the Sasaki-Nakamura equation in the Kerr spacetime \cite{Mino97}
or by deriving from scratch the Regge-Wheeler and Zerilli equations with leading-order rotational corrections~\cite{Pani_2013}.
However,  the eikonal limit of this term is $-2m\omega \Omega + {\cal O} (\ell^{-1})$ which means that for the purpose of this
work it will suffice to assume the same rotational term in both equations.

The system~\eqref{sysWaves} includes as a limiting case the perturbation equations discussed in Paper I.
Those equations describe a massless scalar field without rotation and no couplings through the field derivatives,
i.e. $\mu = \Omega = \gamma_{\psi\Theta} = \delta_{\psi\Theta} =0$.  In addition, the remaining coupling $\beta_{\psi\Theta}$
was assumed to be $\omega$-independent.

The eikonal ansatz
\be
\psi (x) = A(x) e^{iS(x)/\epsilon}, \qquad \Theta (x) = B(x) e^{i H(x)/\epsilon},
\label{eq:ansatz}
\ee
where $\epsilon \ll 1$ is a bookkeeping parameter, leads to
\begin{align}
& A e^{i S/\epsilon}
\left [\, - \frac{(S_{,x})^2}{\epsilon^2}  + \omega^2 -2m\omega \Omega - \ell(\ell + 1) \tilde{V}
 - \frac{6M}{r}  \tilde{V} \right.
\nn \\
& \left. + \frac{i}{\epsilon} \left ( \frac{2  A_{,x}}{A} S_{,x} +  S_{,xx} \right )
+  \frac{A_{,xx}}{A}  \, \right ]  = \beta_\psi B e^{i H/\epsilon},
\label{waveTeik}
\end{align}
and
\begin{align}
& B e^{i H/\epsilon}
\left [\, - \frac{(H_{,x})^2}{\epsilon^2}  + \omega^2 -2m\omega \Omega - \tilde{V} \left \{  \ell (\ell+1)  \alpha + \mu^2 r^2 \right \}
\right.
\nn \\
& \left. + \frac{i}{\epsilon} \left \{ \frac{2 B_{,x}}{B} H_{,x} + g H_{,x} + H_{,xx} \right \}
+ \frac{1}{B} \left ( B_{,xx} + g B_{,x} \right ) \right.
\nn \\
& \left.  - \frac{2 M}{r} \tilde{V} \zeta \, \right ] = A e^{i S/\epsilon}
\left [\,  \beta_\Theta  -\frac{(S_{,x})^2}{\epsilon^2}  b_2 + b_1 A_{,x} + b_2 A_{,xx} \right.
 \nn \\
& \left. - \frac{i}{\epsilon} \left \{ \frac{2 A_{,x}}{A} b_2 S_{,x} + b_1 S_{,x} + b_2 S_{,xx} \right \}  \, \right ],
\label{waveSeik}
\end{align}
where $\tilde{V} (r) = f(r)/r^2$. In the canonical case where the coupling functions vanish at the event horizon and
at infinity, the appropriate QNM boundary conditions for the phase function are,
\begin{align}
& \left \{ \frac{S}{\epsilon},  \frac{H}{\epsilon} \right \} \to  - \sqrt{\omega^2 - 2m \omega \Omega}\, x, \quad\textrm{as}\quad x \to - \infty,
\\
& \frac{S}{\epsilon} \to \omega x, \quad  \frac{H}{\epsilon} \to  \sqrt{\omega^2 - \mu^2}\, x, \quad\,\,\textrm{as}\quad x \to + \infty.
\end{align}

The strategy for manipulating the above system is to solve Eq.~\eqref{waveTeik} for $B$ and insert the result in
Eq.~\eqref{waveSeik}, expecting that the amplitudes and the exponential terms will not appear in the final result
(an alternative method is discussed at the very end of this section).
The eikonal limit is to be taken \emph{afterwards} in this final expression. As already discussed in the eikonal calculation of Paper I,
we assume $\ell \epsilon = m \epsilon = {\cal O} (1)$ and $\omega = {\cal O}(\ell)$ (as first established by Press~\cite{Press:1971wr}).
The scalar mass $\mu$ is an independent parameter without any eikonal scaling. However, we are going to formally treat
it as a leading-order parameter since we would like our model to include scalar fields with $\mu \sim \omega $.
In addition, we work to leading order in $\Omega$. The resulting equation coming out of the system (\ref{waveTeik})-(\ref{waveSeik})
is
\begin{widetext}
\begin{align}
& \omega^4 + \tilde{V} \left ( \ell^2 \alpha + r^2 \mu^2 \right ) \left [  \ell^2 \tilde{V}  + \frac{(S_{,x})^2}{\epsilon^2}  \right ]
-\omega^2 \left [\, \tilde{V} \left \{ \ell^2 (1+\alpha) + \mu^2 r^2 \right \} + \frac{1}{\epsilon^2} \left \{ (S_{,x})^2 + (H_{,x})^2 \right \} \, \right ]
+ \frac{(H_{,x})^2}{\epsilon^2} \ell^2 \tilde{V} + \frac{(S_{,x} H_{,x})^2}{\epsilon^4}
\nn \\
& -\beta_{\psi\Theta} - \frac{1}{A} \left ( A_{,x}\gamma_{\psi\Theta}  + A_{,xx} \delta_{\psi\Theta} \right ) +
\frac{\beta_\psi}{A} e^{i(H-S)/\epsilon}
\left ( g B_{,x} + B_{,xx} \right  ) +  \frac{(S_{,x})^2}{\epsilon^2}\delta_{\psi \Theta}
- \frac{i}{\epsilon^3}  \Big [\, \left ( g H_{,x}  +  H_{,xx} \right )  (S_{,x} )^2
\nn \\
& + \left ( \frac{2 A_{,x}}{A} S_{,x}  + S_{,xx} \right ) (H_{,x})^2\, \Big ] + \frac{1}{\epsilon} \Big [\, -\frac{2i A_{,x}}{A} S_{,x} \tilde{V}
( \ell^2 \alpha + r^2 \mu^2)  - i   \tilde{V} \left \{ g H_{,x} \ell^2 + H_{,xx} \ell^2  + S_{,xx} \left ( \ell^2 \alpha
+ r^2 \mu^2 \right )  \right \} \, \Big ]
\nn \\
& + \frac{1}{\epsilon^2} \Big [\, \ell \tilde{V} \left \{ \alpha (S_{,x})^2 + (H_{,x})^2 \right \} \, \Big ]
+ \omega^2 \left [\, -\ell \tilde{V} (1+\alpha) + \frac{i}{\epsilon} \left ( g H_{,x} +  \frac{2 A_{,x}}{A} S_{,x}
+  H_{,xx} + S_{,xx} \right ) \,\right ]  + \ell \tilde{V}^2 (2 \ell^2 \alpha + r^2 \mu^2 )
\nn \\
& -\frac{i}{\epsilon} \left [\,  S_{,x} \left ( \gamma_{\psi\Theta} + \frac{2 A_{,x}}{A} \delta_{\psi\Theta} \right )
+ S_{,xx} \delta_{\psi\Theta} - \frac{2 B_{,x}}{A} e^{i(H-S)/\epsilon} H_{,x} \beta_\psi \, \right ]
 + 2 m \Omega \omega \Big [\,  \frac{1}{\epsilon^2} \left \{  (S_{,x})^2 + (H_{,x})^2 \right \}  -2\omega^2
 \nn \\
& + \tilde{V} \left \{ \ell^2 (1+\alpha) + \mu^2 r^2 \right  \} \, \Big ]
+ 2m\Omega \left [\, \ell \tilde{V} (1+\alpha) -\frac{i}{\epsilon} \left \{ H_{,xx} + S_{,xx} + g H_{,x} + \frac{2 A_{,x}}{A} S_{,x}  \right \}   \, \right ]
= 0  + {\cal O} \left (\epsilon^{-2} \right).
\label{waveTSeik}
\end{align}
\end{widetext}
This expression displays all ${\cal O}(\epsilon^{-4})$ leading-order and ${\cal O}(\epsilon^{-3})$ subleading-order terms. All ${\cal O} (\epsilon^{-2})$
terms (and higher order) have been omitted as they will not play any role in the subsequent analysis. Notice that all coupling terms have been retained
due to their yet unspecified eikonal order.
Equation~(\ref{waveTSeik}) contains a number of residual coupling terms with the exponential $e^{i(H-S)/\epsilon}$, the amplitudes $A, B$ and their derivatives
appearing in them. These terms are clearly undesirable because they inhibit the proper implementation of the eikonal limit. Fortunately, all these
terms can be effectively removed from the group of leading-order terms by placing suitable constraints on the
eikonal order of the coupling parameters appearing in them. At the same time, we are interested in calculating `canonical' QNMs, which have
a nontrivial GR limit. This requirement simply means that no coupling term in (\ref{waveTSeik}) should exceed ${\cal O}(\epsilon^{-4})$.
More exotic scenarios where $\omega$ is allowed to scale with a higher power of $\ell$ and the coupling terms can exceed ${\cal O}(\epsilon^{-4})$
are discussed in Sec.~\ref{sec:exotic}.

According to this course of action we can remove the term  $\sim \epsilon^{-1} e^{i(H-S)/\epsilon} (B_{,x}/A) S_{,x} \beta_\psi $ by assuming
$\beta_\psi \leq {\cal O} (\ell^3)$. This scaling automatically pushes `under the carpet' the term $\sim \beta_\psi e^{i(H-S)/\epsilon} ( g B_{,x} + B_{,xx} ) $.
For the same reason the removal of the terms $\sim \epsilon^{-2} (S_{,x})^2 \delta_{\psi \Theta}$ and  $\sim \epsilon^{-1} S_{,x} \gamma_{\psi \Theta}$
is accompanied, respectively, by the constraints $ \delta_{\psi\Theta} \leq {\cal O} (\ell^2)$ and $\gamma_{\psi\Theta} \leq {\cal O} (\ell^3)$.
These constraints will be subject to revision once we move to the subleading-order analysis.

Before proceeding with our eikonal analysis we should emphasize that another approach to the solution of the system
(\ref{waveTeik})-(\ref{waveSeik}) consists in the elimination of $B$ together with its derivatives $B_{,x}, B_{,xx}$ in the latter equation.
This equivalent method leads to the same leading-order results as the ones we discuss in the following section. The agreement extends to the
subleading-order results provided $S_{,xx} = H_{,xx} $ at the same point where the eikonal formulae are to be evaluated
(this is the `potential peak'  $r=r_{\rm m}$ where $S_{,x} = H_{,x} =0$, see below for details).

%%%%%%%%%%%%%%%%%%%%%%%%%%%%%%%%%%%%%%%%%%%%%%%%%%%%%%%%%

\section{Leading-order eikonal analysis}
\label{sec:leading_order}

Having  at our disposal the general expression (\ref{waveTSeik}) it is straightforward to isolate the leading-order terms:
\begin{align}
& \omega^4 -\omega^2 \left [\, \tilde{V} \left \{ \ell^2 (1+\alpha) + \mu^2 r^2 \right \} + \frac{1}{\epsilon^2} \left \{ (S_{,x})^2 + (H_{,x})^2 \right \} \, \right ]
\nn \\
& + \tilde{V} \left ( \ell^2 \alpha + r^2 \mu^2 \right ) \left [  \ell^2 \tilde{V}  + \frac{(S_{,x})^2}{\epsilon^2}  \right ]
+ \frac{(H_{,x})^2}{\epsilon^2} \ell^2 \tilde{V}
\nn \\
&  + \frac{(S_{,x} H_{,x})^2}{\epsilon^4}  -\beta_{\psi\Theta}^{(0)}  +  \frac{(S_{,x})^2}{\epsilon^2}\delta_{\psi \Theta}^{(0)}
-\frac{i}{\epsilon} \left [\,  S_{,x}  \gamma_{\psi\Theta}^{(0)} \right.
\nn \\
& \left. - \frac{2 B_{,x}}{A} e^{i(H-S)/\epsilon} H_{,x} \beta_\psi^{(0)} \, \right ]
+ 2 m \Omega \omega \left [\,  \frac{1}{\epsilon^2} \left \{  (S_{,x})^2 + (H_{,x})^2 \right \}  \right.
\nn \\
&\left.  -2\omega^2 + \tilde{V} \left \{ \ell^2 (1+\alpha) + \mu^2 r^2 \right  \} \, \right ] = 0,
\label{leadEq1}
\end{align}
where we have assumed the following eikonal expansions for the coupling parameters,
\begin{align}
\beta_{\psi\Theta} &= \underbrace{\beta_{\psi\Theta}^{(0)}}_{\order4} + \underbrace{\beta_{\psi\Theta}^{(1)}}_{\order3} + \, {\cal O}(\ell^{2}),
\label{coupling_scaling1a}
\\
 \delta_{\psi\Theta} &= \underbrace{\delta_{\psi\Theta}^{(0)}}_{\order2} + \underbrace{\delta_{\psi\Theta}^{(1)}}_{{\cal O}(\ell)} + \, {\cal O}(1),
\label{coupling_scaling1b}
\end{align}
and
\begin{align}
\beta_\psi &=  \underbrace{\beta_\psi^{(0)}}_{\order3} + \underbrace{\beta_\psi^{(1)}}_{\order2} +  \, {\cal O}(\ell),
\\
\gamma_{\psi\Theta} & =  \underbrace{\gamma_{\psi\Theta}^{(0)}}_{\order3} + \underbrace{\gamma_{\psi\Theta}^{(1)}}_{\order2} +  {\cal O}(\ell),
\label{coupling_scaling2}
\end{align}
with the underbraces (here and elsewhere in the text) indicating the eikonal order of the different terms.
We note that the $\ell$-scaling of the last two parameters  will be reduced by one order as a consequence of the constraints placed by
Eq.~(\ref{Eqrm}), see the following section.

The next step is to assume a  `peak' radius $r=r_{\rm m}$ where $S_{,x} = H_{,x}= 0$. Then Eq. (\ref{leadEq1}) yields
\begin{align}
& \omega^4 -\omega^2 \tilde{V}_{\rm m} \left [ \ell^2 (1+\alpha_{\rm m}) + \mu^2 r^2_{\rm m} \right ]
 + \ell^2 \tilde{V}^2_{\rm m} \left ( \ell^2 \alpha_{\rm m} +  \mu^2 r^2_{\rm m} \right )
 \nn \\
& -(\beta_{\psi\Theta}^{(0)})_{\rm m} + 2 m \Omega_{\rm m} \omega \Big [\,  \tilde{V}_{\rm m} \left \{ \ell^2 (1+\alpha_{\rm m})
+ \mu^2 r^2_{\rm m} \right  \}   -2\omega^2 \, \Big ]
\nn \\
&= 0.
\label{leadEq3}
\end{align}
We observe that the dominant eikonal coupling terms $\beta_{\psi}^{(0)}$, $\gamma_{\psi\Theta}^{(0)}$
and $\delta_{\psi\Theta}^{(0)}$ do not contribute to the leading-order frequency $\omega$.
In general $\beta_{\psi\Theta}$ may be a function of $\omega$. To account for such possibility we assume that the
leading-order eikonal term can be written as
\be
\beta_{\psi\Theta}^{(0)} (r) =\beta_4 (r) \ell^4 + \beta_3 (r) \ell^3 \omega + \beta_2(r) \ell^2 \omega^2.
\ee
This expansion implicitly assumes that all $\omega$-dependence arises from time derivatives of
the perturbation variables and that these derivatives are  of second order at most hence the
absence of a $\sim \ell \omega^{3}$ term. With this expansion, Eq.~(\ref{leadEq3}) becomes
\begin{align}
&\omega^4 - \omega^2 \left [\, \tilde{V}_{\rm m} \{  \ell^2 (1+ \alpha_{\rm m} )  +  \mu^2 r^2_{\rm m} \}
+ \ell^2  \beta_{\rm 2m} \, \right ]
\nn \\
& - \ell^3 \beta_{\rm 3m} \omega + \ell^2 \left [\, \tilde{V}_{\rm m}^2
(  \ell^2 \alpha_{\rm m}  +  \mu^2 r^2_{\rm m} ) - \ell^2 \beta_{\rm 4m} \, \right ]
\nn \\
& + 2 m \Omega_{\rm m} \omega \Big [\,  \tilde{V}_{\rm m} \left \{ \ell^2 (1+\alpha_{\rm m})
+ \mu^2 r^2_{\rm m} \right  \}   -2\omega^2 \, \Big ] = 0.
\label{leadEq4}
\end{align}
Let us first consider the nonrotating limit ($\Omega=0$). Although Eq.~\eqref{leadEq4}
can be solved analytically in this limit, the resulting roots are too cumbersome. Instead, we focus on the simpler
case $\beta_3 =0$ for which (\ref{leadEq4}) is a biquadratic with roots,
\begin{align}
\omega_\pm^2  &= \frac{1}{2} \ell^2 \beta_{\rm 2m} + \frac{1}{2} \tilde{V}_{\rm m} \left [ \ell^2 (1+ \alpha_{\rm m} ) + \mu^2 r^2_{\rm m}\right ]
\nn \\
&\quad \pm \left\{\left [ \tilde{V}_{\rm m} \{  \ell^2 (1+\alpha_{\rm m} ) +  \mu^2 r^2_{\rm m}  \}  +\ell^2 \beta_{\rm 2m}
     \right ]^2 \right.
\nn \\
&\quad \left. -4  \ell^2 \left [  \tilde{V}^2_{\rm m} ( \ell^2  \alpha_{\rm m} +  \mu^2 r^2_{\rm m} )  -\ell^2\beta_{\rm 4m}
\right ] \right\}^{1/2}.
\label{ompm}
\end{align}
Switching back to the system with rotation, we use in (\ref{leadEq4}) the following expansion for the frequency,
\be
\omega = \sigma_\pm = \omega_\pm +  m \Omega_{\rm m} \omega_1.
\label{eq:def_omega_leading_order}
\ee
The rotational correction $\omega_1$ is easily obtained from the ${\cal O} (\Omega)$ part of (\ref{leadEq4}),
\begin{align}
\omega_{1} & = \frac{  \tilde{V}_{\rm m} [ \ell^2 (1+  \alpha_{\rm m}) +  \mu^2 r^2_{\rm m} ] - 2\omega^2_\pm}
{\tilde{V}_{\rm m} \left [\, \ell^2 (1+\alpha_{\rm m}) + \mu^2 r^2_{\rm m}\, \right ]  + \ell^2 \beta_{2m} -2\omega^2_\pm }
\nn \\
& =  1 \pm \ell^2 \beta_{\rm 2m}
\left  \{ \left [ \tilde{V}_{\rm m} \{  \ell^2 (1+\alpha_{\rm m} ) +  \mu^2 r^2_{\rm m}  \}  +\ell^2 \beta_{\rm 2m} \right ]^2 \right.
\nn \\
& \left. \quad -\, 4\ell^2 \left [  \tilde{V}^2_{\rm m} ( \ell^2  \alpha_{\rm m} +  \mu^2 r^2_{\rm m} )  -\ell^2\beta_{\rm 4m}  \right ]  \right \}^{-1/2},
\label{om1root1}
\end{align}
where the second equation follows from using (\ref{ompm}).

In order to obtain a fully consistent ${\cal O} (\Omega)$ result for $\sigma_\pm$ we need to account for the
rotational correction in $r_{\rm m}$ itself. Decomposing $r_{\rm m}$ as
\be
r_{\rm m} = r_0 + r_1, \qquad r_1 = {\cal O} (\Omega),
\ee
we ought to Taylor-expand  the $\omega_\pm (r_{\rm m})$ roots in Eq.~(\ref{ompm}):
\be
\omega_\pm  = \omega_{0}  + \omega_{0}^\p r_1 + {\cal O}(\Omega^2), \qquad \omega_{0}  \equiv \omega_\pm (r_0).
\ee
Then,
\be
\sigma_\pm =  \omega_{0}  + \omega_{0}^\p r_1 + m \Omega_0 \omega_1,
\ee
where $\Omega_0 = \Omega (r_0)$ and $\omega_1$ is given by (\ref{om1root1}) with $r_{\rm m} \to r_0$.

%%%%%%%%%%%%%%%%%%%%%%%%%%%%%%%%%%%%%%%%%%%%%%

\section{The effective potential}
\label{sec:Veff}

To complete our leading-order analysis we need to derive an equation for $r_{\rm m}$, to enable us to calculate
$\sigma_{\pm}$.
To do so, we first take an $r$-derivative of (\ref{leadEq1}) and then set $r = r_{\rm m}$ (where $S_{,x}=H_{,x}=0$), which yields
\begin{align}
& -\omega^2 \left [\, \tilde{V}_{\rm m} \left ( \ell^2 \alpha^\p_{\rm m} + 2\mu^2 r_{\rm m} \right ) + \tilde{V}^\p_{\rm m}
\left \{ \ell^2 (1+\alpha_{\rm m} ) + \mu^2 r^2_{\rm m} \right \} \, \right]
\nn \\
& + \ell^2 \tilde{V}^2_{\rm m} \left (\ell^2 \alpha^\p_{\rm m} + 2 \mu^2 r_{\rm m} \right )
 + ( \tilde{V}^2 )^\p_{\rm m} \ell^2 \left ( \ell^2 \alpha_{\rm m} + \mu^2 r^2_{\rm m} \right )
\nn \\
& -( \beta^{(0)}_{\psi\Theta})_{\rm m}^\p - \frac{ix^\p_{\rm m}}{\epsilon} \left ( S_{,xx}  \gamma_{\psi\Theta}^{(0)}
- \frac{2 B_{,x}}{A} H_{,xx}   \beta_\psi^{(0)} e^{i(H-S)/\epsilon} \right )_{\rm m}
\nn \\
& + 2m \omega \left [\, -2 \Omega^\p_{\rm m} \omega^2 + \Omega_{\rm m} \tilde{V}_{\rm m} (\ell^2 \alpha_{\rm m}^\p + 2 \mu^2 r_{\rm m })
\right.
\nn \\
& \left. +\, ( \Omega \tilde{V} )^\p_{\rm m}  \left \{ \ell^2 (1+\alpha_{\rm m} ) + \mu^2 r_{\rm m}^2 \right \} \, \right ] = 0,
\label{EqrmFull}
\end{align}
where $x^\p = dx/dr$. The third line's parentheses term is clearly undesirable and can only be removed by reducing by one order
the eikonal scaling of the coupling parameters, such that hereafter
\be
\{ \beta_\psi^{(0)}, \gamma_{\psi\Theta}^{(0)} \} = \order{2}.
\ee
By doing so we are also dropping terms proportional to $\{B_{,x},\, H_{,x}\}$ in the calculation of the previous
section. This has a harmless effect because these terms did not contribute to the final result.

After the removal of the parentheses term, Eq.~\eqref{EqrmFull} can be identified as the radial derivative of an effective potential,
\be
U_{\rm eff}^\p (r_{\rm m}, \omega)  = 0,
\label{Eqrm}
\ee
defined as
\be
U_{\rm eff} (r,\omega) =  V_{\rm eff} (r, \omega) + V_{\rm \Omega} (r, \omega),
\label{Ueff}
\ee
which we decomposed into nonrotating and rotating contributions, defined respectively as
\begin{align}
 V_{\rm eff} (r, \omega) & =  \omega^2  \tilde{V} \left [ \ell^2 (1+\alpha) + \mu^2 r^2  \right]
 \label{eq:v_eff}
\nn \\
& \quad - \ell^2\tilde{V}^2  \left ( \ell^2 \alpha + \mu^2 r^2 \right ) + \beta_{\psi\Theta}^{(0)},
\\
 V_{\rm \Omega} (r, \omega) &= 2m \omega \Omega \left [  2 \omega^2  -\tilde{V}   \left \{ \ell^2 (1+\alpha ) + \mu^2 r^2 \right \}  \right ].
\end{align}
The peak $r_{\rm m} = r_0$ of the nonrotating system can be associated with an extremum of the potential $V_{\rm eff}$, i.e.
\be
V_{\rm eff}^\p  ( r_0, \omega_0 )= 0.
\label{eq:def_r_0}
\ee
The rotational correction $r_1$ can be obtained by expanding $U_{\rm eff}^\p (r_{\rm m}, \sigma_\pm ) =0$
around $( r_0, \omega_0 ) $. We obtain,
\be
[ V_{\rm eff}^{\pp} \, r_1 +V_{{\rm eff},\omega}^\p  ( \omega_0^\p r_1 + m \Omega \omega_1 )
+ V_{\rm \Omega}^\p ]_0 = 0,
\ee
and
\be
r_1 = - \left ( \frac{V_{\rm \Omega}^\p   +  m \Omega  \omega_1 V_{{\rm eff},\omega}^\p }
{ V_{\rm eff}^{\pp} + \omega_0^\p V_{{\rm eff},\omega}^\p} \right )_0,
\label{eq:def_r_1}
\ee
where the subscript `0' means that the functions inside the brackets should be evaluated at $(r_0,\omega_0)$.
This completes our leading-order eikonal analysis of Eqs.~\eqref{sysWaves}.

%%%%%%%%%%%%%%%%%%%%%%%%%%%%%%%%%%%%%%%%%%%%%%%%%%%%

\section{Subleading-order eikonal analysis}
\label{sec:subleading_order}

After obtaining the leading-order eikonal formulae we can now proceed to the subleading-order calculation.
As in Paper I, carrying the analysis to this order will yield formulae for the imaginary part and the subleading
real part of the QNM frequencies.

The first step of the subleading analysis is the frequency expansion,
\begin{align}
\omega &= \omega_R + i \omega_I = \underbrace{\omega_R^{(0)}}_{{\cal O}(\ell)}
+ \underbrace{ \omega_R^{(1)} + i \omega_I }_{{\cal O} (1)} + {\cal O} (\ell^{-1}), \quad
\label{om_scaling}
\\
\omega_R^{(0)} &= \sigma_\pm.
\label{eq:def_omega_R_zero}
\end{align}
The coupling functions are still expanded as in \eqref{coupling_scaling1a}-\eqref{coupling_scaling2}
\begin{subequations}
\begin{align}
\beta_\psi &=  \underbrace{\beta_\psi^{(0)}}_{\order2} + \underbrace{\beta_\psi^{(1)}}_{{\cal O}(\ell)} +\,{\cal O}(1), \quad
\\
\gamma_{\psi\Theta} &=  \underbrace{\gamma_{\psi\Theta}^{(0)}}_{\order2} + \underbrace{\gamma_{\psi\Theta}^{(1)}}_{{\cal O}(\ell)} +\,{\cal O}(1).
\end{align}
\label{coupling_scaling3}
\end{subequations}

The subleading equation corresponds to the ${\cal O} (\epsilon^{-3})$ part of (\ref{waveTSeik})
after the coupling parameters expansions have been inserted. When evaluated at $r=r_{\rm m}$ that equation becomes,
\begin{widetext}
\begin{align}
& - (\beta_{\psi\Theta}^{(1)})_{\rm m} + \ell \tilde{V}_{\rm m}^2 \left [\, 2\ell^{2} \alpha_{\rm m} + \mu^2  r^2_{\rm m} \, \right ]
-  \ell \tilde{V}_{\rm m} (1+ \alpha_{\rm m} ) (\omega_R^{(0)})^2
+ 2 \omega_R^{(0)} \omega_R^{(1)} \left [\,  2 (\omega_R^{(0)} )^2 -  \tilde{V}_{\rm m} \left \{ \, \ell^2 (1+ \alpha_{\rm m}) + r^2_{\rm m} \mu^2
\,\right \}   \, \right ]
\nn \\
& + i \Big [ \,  2 \omega_R^{(0)} \omega_I \left \{\,  2 (\omega_R^{(0)} )^2  -\tilde{V}_{\rm m} \left [ \, \ell^2 (1+ \alpha_{\rm m}) + r^2_{\rm m} \mu^2
\,\right ] \, \right \} + (S_{,xx})_{\rm m} \left \{  (\omega_R^{(0)} )^2 -  \tilde{V}_{\rm m} \left (\ell^2 \alpha_{\rm m} + r^2_{\rm m} \mu^2 \right )
- (\delta^{(0)}_{\psi\Theta})_{\rm m}  \right \}
\nn \\
& + (H_{,xx})_{\rm m} \left \{  (\omega_R^{(0)} )^2 -\ell^2 \tilde{V}_{\rm m} \right \}  \, \Big ]
+ 2m \Omega_{\rm m} \left [\, \ell \tilde{V}_{\rm m} (1+\alpha_{\rm m} ) \omega_R^{(0)}
+ \omega_R^{(1)} \left \{ \tilde{V}_{\rm m} [ \ell^2 (1+\alpha_{\rm m}) + \mu^2 r^2_{\rm m}]  \right.\right.
\nn \\
& \left. \left. - 6 (\omega_R^{(0)})^2 \right \} \, \right ] + 2 i m\Omega_{\rm m} \left [\, \omega_I \left \{ \tilde{V}_{\rm m}  [\ell^2 (1+\alpha_{\rm m} )
+ \mu^2 r^2_{\rm m}] - 6 (\omega_R^{(0)})^2 \right \} - \omega_R^{(0)} \left \{ (S_{,xx})_{\rm m} + (H_{,xx})_{\rm m}  \right \} \, \right ]= 0.
\label{subleadEq1}
\end{align}
\end{widetext}
Equation~\eqref{subleadEq1} has two features worth noticing. First, the nonrotating part depends only on $ \{ \,\beta_{\psi\Theta}^{(1)},
\delta_{\psi\Theta}^{(0)}, \mu^2 \, \}$ while $\{\, \beta_{\psi}^{(0)}$, $\gamma_{\psi\Theta}^{(0)}\, \}$ are absent altogether.
Second, the rotating part depends only on $\mu^{2}$, but implicitly on the $\beta_{\psi\Theta}^{(0)}$ through $\omega_{R}^{(0)}$.  The absence of
$\gamma_{\psi\Theta}^{(0)}$ means that the coupling term $b_1 (d\psi/dx$) does not contribute to the eikonal QNM frequency [cf.~\eqref{defcouplings}].

As in Paper I, the imaginary (real) part of (\ref{subleadEq1}) will furnish $\omega_I$ ($\omega_R^{(1)}$). However, the identification of these two parts
requires some prior input for the $\omega$-dependence of $\beta_{\psi\Theta}^{(1)}$ and $\delta^{(0)}_{\psi\Theta}$.
We first consider the simplest scenario in which these two coupling parameters are real and frequency independent.

%%%%%%%%%%%%%%%%%%%%%%%%%%%%%%%%%%%%%%%%%%%%%%%%%%%%%%%%%%%%%%

\subsection{Subleading-order analysis with $\omega$-independent and real $\beta_{\psi\Theta}^{(1)}$ and $\delta^{(0)}_{\psi\Theta}$}
\label{sec:sublead_no_omega}

Isolating the real and imaginary parts of \eqref{subleadEq1} we find respectively
\begin{align}
& 2 \omega_R^{(0)} \omega_R^{(1)}   \left [\,  2 (\omega_R^{(0)} )^2 -  \tilde{V}_{\rm m} \left \{ \, \ell^2 (1+ \alpha_{\rm m}) + \mu^2 r^2_{\rm m}
\,\right \}   \, \right ]
\nn \\
&-  \ell \tilde{V}_{\rm m}  (1+ \alpha_{\rm m} )   (\omega_R^{(0)})^2
- (\beta_{\psi\Theta}^{(1)})_{\rm m}
&
\nn \\
&+ \ell \tilde{V}_{\rm m}^2 (\, 2\ell^2 \alpha_{\rm m} + \mu^2  r^2_{\rm m} \, )
+ 2m \Omega_{\rm m} \Big [\, \ell \tilde{V}_{\rm m}  (1+\alpha_{\rm m} )\, \omega_R^{(0)}
\nn \\
&+ \omega_R^{(1)} \left \{ \tilde{V}_{\rm m} [ \ell^2 (1+\alpha_{\rm m}) + \mu^2 r^2_{\rm m}]  - 6 (\omega_R^{(0)})^2 \right \} \, \Big ]
= 0,
\label{sublead_EqomR1}
\end{align}
and
\begin{align}
&  2 \omega_R^{(0)} \omega_I \left [\,  2 (\omega_R^{(0)} )^2  -\tilde{V}_{\rm m} \left \{ \, \ell^2 (1+ \alpha_{\rm m}) + \mu^2 r^2_{\rm m}
\,\right \} \, \right ]
\nn \\
&  + (S_{,xx})_{\rm m} \left [  (\omega_R^{(0)} )^2 -  \tilde{V}_{\rm m} \left (\ell^2 \alpha_{\rm m} + \mu^2 r^2_{\rm m} \right )
- (\delta^{(0)}_{\psi\Theta})_{\rm m}  \right ]
\nn \\
& + (H_{,xx})_{\rm m} \left [  (\omega_R^{(0)} )^2 -\ell^2 \tilde{V}_{\rm m} \right ]
\nn \\
& + 2  m\Omega_{\rm m} \Big [\, \omega_I \left \{ \tilde{V}_{\rm m}  [\ell^2 (1+\alpha_{\rm m} ) + \mu^2 r^2_{\rm m}]
- 6 (\omega_R^{(0)})^2 \right \}
\nn \\
& - \omega_R^{(0)} \left \{ (S_{,xx})_{\rm m} + (H_{,xx})_{\rm m}  \right \} \, \Big ]= 0.
\label{sublead_EqomI}
\end{align}
This pair of equations can be algebraically solved for $\omega_R^{(1)}$ and $\omega_I$. Next, we substitute
[cf. Eqs.~\eqref{eq:def_omega_leading_order} and~\eqref{eq:def_omega_R_zero}]
\be
\omega_R^{(0)} = \sigma_{\pm} = \omega_\pm + m \Omega_{\rm m} \omega_1,
\label{eq:omega_r_0_sigma}
\ee
and then expand to leading order in rotation. These steps lead to
\begin{align}
\omega_I &= - \frac{1}{2\omega_\pm  \cW_{\rm m}} \left \{\, (H_{,xx})_{\rm m} \left [\, \omega^2_\pm  -\ell^2 \tilde{V}_{\rm m} \,\right ]
\right.
\nn \\
&\quad \left. + (S_{,xx})_{\rm m} \left [\, \omega^2_\pm - \tilde{V}_{\rm m} (\ell^2 \alpha_m + \mu^2 r^2_{\rm m} )
-(\delta_{\psi\Theta}^{(0)})_{\rm m} \,\right ] \, \right \}
\nn \\
&\quad + \frac{2 m \Omega_{\rm m}}{(2 \omega_\pm \cW_{\rm m})^2 }  (\omega_1 -1) \left [\, (H_{,xx})_{\rm m} \cC_H  +  (S_{,xx})_{\rm m} \cC_S \, \right ],
\nn \\
\label{eq:omIroot1}
\\
\omega_R^{(1)} &= \frac{\cB_{\rM}}{2 \omega_{\pm} \cW_{\rM}} + \frac{2 m \Omega_{\rM}}{(2\omega_{\pm} \cW_\rM)^2}(\omega_1 - 1) \cE_{\rM} ,
\label{eq:omR1root1}
\end{align}
where we defined the auxiliary functions
\begin{align}
\cC_H &= \omega^2_\pm \left  [\,  2\omega_\pm^2
+ \tilde{V}_{\rm m} \{ \ell^2 (\alpha_{\rm m} -5) + \mu^2 r^2_{\rm m} \}\, \right ]
\nn \\
&\quad + \ell^2 \tilde{V}^2_{\rm m}  \left [\,  \ell^2 (1+ \alpha_{\rm m} ) + \mu^2 r^2_{\rm m}\, \right ],
\\
\cC_S &= \omega_\pm^2 \left [\, 2\omega_\pm^2 + \tilde{V}_{\rm m} \left \{\, \ell^2 (1-5 \alpha_{\rm m}) - 5 \mu^2 r^2_{\rm m}  \, \right \}
\right.
\nn \\
&\quad\left. -\, 6 (\delta_{\psi\Theta}^{(0)})_{\rm m} \,\right ]
+ \tilde{V}_{\rm m}   \left [\,  \ell^2 (1+ \alpha_{\rm m} ) + \mu^2 r^2_{\rm m}\, \right ]
\nn \\
& \quad \times  \left [\,\tilde{V}_{\rm m} (  \ell^2 \alpha_{\rm m}  + \mu^2 r^2_{\rm m} ) + (\delta_{\psi\Theta}^{(0)})_{\rm m} \, \right ],
\\
\cW &=  2 \omega^2_\pm
-\tilde{V} \left [ \ell^2 (1+ \alpha) + \mu^2 r^2 \right ] \,,
\label{calWdef}
\end{align}
and
\begin{align}
\cB &= \ell \tilde{V} \left[ \omega_{\pm}^2 (1 + \alpha) - \tilde{V}(2 \ell^2 \alpha + \mu^2 r^2) \right] + (\beta_{\psi\Theta})^{(1)}_\rM,
\\
\cE &= - \ell \vt^3 (2 \ell^2 \alpha + \mu^2 r^2)[\ell^2(1+\alpha) + \mu^2 r^2]
\nn \\
&\quad - \ell \vt^2 \omega_{\pm}^2 \left[ \ell^2(\alpha^2 - 10\alpha + 1) + (\alpha - 5) \mu^2 r^2 \right]
\nn \\
&\quad + \vt \left[\beta_{\psi\Theta}^{(1)}\{ \ell^2(1+\alpha) + \mu^2 r^2 \} - 2 \ell (1 + \alpha)\, \omega_{\pm}^4 \right].
\nn \\
\end{align}

To proceed we must calculate the derivatives $(S_{,xx})_{\rm m}$ and $(H_{,xx})_{\rm m}$, which
appear in Eq.~\eqref{eq:omIroot1}.
To do so, we first write Eq.~\eqref{leadEq1} in the following equivalent form (after taking
into account the revised $\ell$-scaling of the coupling parameters):
\begin{align}
&\omega^4 -U_{\rm eff} (r,\omega)   + \frac{(S_{,x})^2}{\epsilon^2}  \left [\, \tilde{V} (\ell^2 \alpha + \mu^2 r^2) + \delta_{\psi\Theta}^{(0)} -\omega^2 \, \right ]
\nn \\
& -\frac{(H_{,x})^2}{\epsilon^2} \left (\,  \omega^2 - \ell^2 \tilde{V} \right ) + \frac{(S_{,x} H_{,x})^2}{\epsilon^4}
+ \frac{2}{\epsilon^2} m \Omega \omega  \left [\,  (S_{,x})^2 \right.
\nn \\
& \left. +\, (H_{,x})^2 \, \right ]  = 0.
\end{align}
This expression can be subsequently Taylor-expanded around $r=r_{\rm m}$ with the help of,
\begin{align}
& S_{,x} (r) \approx x^\p_{\rm m} (S_{,xx})_{\rm m} \delta r, \quad H_{,x} (r) \approx x^\p_{\rm m} (H_{,xx})_{\rm m}\delta r,
\nn \\
& U_{\rm eff} (r,\omega)  \approx \omega^4 + \frac{1}{2} U_{\rm eff}^{\pp}  (r_{\rm m}, \omega) \delta r^2,
\end{align}
with $\delta r \equiv r -r_{\rm m}$. The resulting ${\cal O} ( \delta r^2 )$ accurate expression with $\omega = \sigma_\pm$ is,
\begin{align}
& - \frac{1}{2} U_{\rm eff}^{\pp}  (r_{\rm m}, \sigma_\pm) + \frac{ ( x^\p_{\rm m} )^2 }{\epsilon^2} \left [\,  (S_{,xx})^2_{\rm m}
 \left \{\, \tilde{V}_{\rm m} (\ell^2 \alpha_{\rm m} + \mu^2 r^2_{\rm m}) \right. \right.
 \nn \\
& \left. \left. + ( \delta_{\psi\Theta}^{(0)})_{\rm m} -\sigma^2_\pm \, \right \} -(H_{,xx})^2_{\rm m} \left (\,  \sigma^2_\pm - \ell^2 \tilde{V}_{\rm m} \right )  \right ]
 \nn \\
 & + \frac{2m}{\epsilon^2} \Omega_{\rm m} \omega_\pm ( x^\p_{\rm m} )^2  \left [\, (S_{,xx})^2_{\rm m} + (H_{,xx})^2_{\rm m}  \, \right ]  = 0.
 \label{eq:SxxIntermediate}
\end{align}
At this point we need to provide a relation between $\ddSm$ and $(H_{,xx})_{\rm m}$. As we did in Paper I, we set
\be
\ddSm =(H_{,xx})_{\rm m}.
\ee
The same condition is enforced by the alternative method discussed at the end of Sec.~\ref{sec:coupledEqs}.
We can subsequently solve  Eq.~\eqref{eq:SxxIntermediate} for $(S_{,xx})^2_{\rm m}$ and then make a slow-rotation expansion to find
\begin{align}
 \frac{(S_{,xx})^2_{\rm m}}{\epsilon^2}  &=  -\left ( \frac{dr}{dx} \right )_{\rm m}^2 \cD_{\rm m}^{-1}
 \left [\, \frac{1}{2} (V_{\rm eff}^{\pp})_{\rm m}   + m \omega_\pm \Omega^{\pp}_{\rm m} \cW_{\rm m} \right.
\nn \\
&\quad \left.-\,2 m \omega_\pm \Omega^\p_{\rm m} (\tilde{V} \left \{ \ell^2 (1+\alpha) + \mu^2 r^2\right \})^\p_{\rm m}
\right.
\nn \\
&\quad \left. -m \omega_\pm \Omega_{\rm m} (\omega_1 -1) \frac{ \cA_{\rm m}}{ \cD_{\rm m}}   \, \right ]\,,
\label{Sxx1}
\end{align}
where $(V_{\rm eff}^{\pp})_{\rm m} \equiv V_{\rm eff}^{\pp} (r_{\rm m}, \omega_\pm)$ and
\begin{align}
\cD &= 2 \omega^2_\pm  - \tilde{V}  \left [\, \ell^2 (1 + \alpha ) + \mu^2 r^2 \, \right ] - \delta_{\psi\Theta}^{(0)}\,,
\label{calDdef}
\\
\cA  &= \tilde{V}^2 ( \ell^2 \alpha^{\pp}  +2 \mu^2) \left [\ell^2 (\alpha -1) + \mu^2 r^2 \right ]
\nn \\
&\quad + 2 \left [\,  \beta_{\psi\Theta}^{(0) \pp } - 2(\tilde{V}^\p)^2 \ell^2 \left ( \ell^2 \alpha + \mu^2 r^2 \right ) \right ]
\nn \\
&\quad +\delta_{\psi\Theta}^{(0)}  ( \tilde{V} [\ell^2 (1+ \alpha ) + \mu^2 r^2 ] )^{\pp}
\nn \\
&\quad + \tilde{V} \Big [\,  2 \tilde{V}^\p  \left (\ell^2 \alpha^\p  + 2 \mu^2 r \right ) \left \{ \ell^2  (\alpha - 3) + \mu^2 r^2 \right \}
\nn \\
&\quad + \tilde{V}^{\pp} \left \{ \ell^2  (\alpha -1) + \mu^2 r^2 \right \}^{2}  \,\Big ] \,.
\label{calAdef}
\end{align}
We take the square root so that $(S_{,xx})_{\rm m} >0$ and once again make a slow-rotation expansion to obtain,
\begin{align}
    \frac{(S_{,xx})_{\rm m}}{\epsilon} = \frac{ \left (dr/dx\right )_{\rm m} }{(2\cD_{\rm m})^{1/2} } | (V_{\rm eff}^{\pp})_{\rm m} |^{1/2}
    \left [ 1 - \frac{{\cal H}_{\rm m}}{V_{\rm eff}^{\pp} (r_{\rm m}, \omega_\pm)}  \right ],
    \nn \\
\label{Sxx2}
\end{align}
where
\begin{align}
\cH &=  2 m \omega_\pm \Omega^\p
 (\tilde{V} \left \{ \ell^2 (1+\alpha) + \mu^2 r^2\right \})^\p
 \nn \\
&\quad -m \omega_\pm \Omega^{\pp} \cW
+ m \omega_\pm \Omega (\omega_1 -1) \frac{\cA}{\cD}.
\end{align}
We can now return to Eq.~(\ref{sublead_EqomI}) and solve for $\omega_I$. We obtain
\be
\omega_I =  - \frac{(S_{,xx})_{\rm m}}{2\omega_\pm  \cW_{\rm m}} \left [\, \cD_{\rm m}
-  \frac{m \Omega_{\rm m}}{\omega_\pm \cW_{\rm m}}  (\omega_1 -1) \, \cC_{\rm m}\, \right ],
\label{omIroot3}
\ee
with
\begin{align}
\cC_{\rm m}
&\equiv \cC_H  +  \cC_S
\nn \\
&= -\, 6\, \omega_\pm^2 \, \delta_{\psi\Theta}^{(0)}
+ 4 \beta_{\psi\Theta}^{(0)}
% \nn \\
+ \tilde{V}^2  \left [ \ell^2 (\alpha -1) + \mu^2 r^2  \right ]^2
\nn \\
% &\quad \times \left [ \ell^2 (\alpha - 1) + \mu^2 r^2  \right ]
  &\quad + \delta_{\psi\Theta}^{(0)}  \tilde{V} \left [ \ell^2 (1+ \alpha ) + \mu^2 r^2  \right ].
\label{calCdef}
\end{align}
To obtain $\cC_{\rm m}$ we used $\omega^4_\pm = V_{\rm eff} (r_{\rm m}, \omega_\pm) + {\cal O} (\Omega)$,
where $V_{\rm eff}$ is given by Eq.~\eqref{eq:v_eff}.

Combining Eqs.~\eqref{omIroot3} and (\ref{Sxx2}) leads to
\begin{align}
\omega_I &=  - \frac{ \left (dr/dx\right )_{\rm m} }{2 \sqrt{2}\omega_\pm  } \frac{|(V_{\rm eff}^{\pp})_{\rm m} |^{1/2}\,\cD_{\rm m}^{1/2}}{\cW_{\rm m} }
\nn \\
&\quad \times \left[ 1 - \frac{\cH_{\rm m}}{(V_{\rm eff}^{\pp})_{\rm m}} - \frac{m \Omega_{\rm m}}{\omega_{\pm} \cW_{\rm m} \cD_{\rm m}} (\omega_1 - 1) \,\cC_{\rm m}\right].
\label{omIroot4}
\end{align}
Our final task for this subleading-order analysis is to introduce $r_{\rm m} = r_0 + r_1$ in (\ref{omIroot4}) and expand around $r_0$:
\begin{align}
\omega_I &=  \cF_0   \left[ 1  -  \frac{\cH_0}{V_{\rm eff}^{\pp}(r_0,\omega_0)}
- \frac{m \Omega_0  \cC_0}{\omega_0 \cW_0 {\cD}_0 } (\omega_1 -1) \, \right] + \cF^\p_0 \, r_1,
\label{eq:omegaI_omega_indep}
\end{align}
where we have defined the function [here $\omega_\pm = \omega_\pm (r_{\rm m})$]
\be
\cF (r_{\rm m}) = -  \frac{ \left (dr/dx\right )_{\rm m}}{2 \sqrt{2} \,\omega_\pm}
\frac{ | (V_{\rm eff}^{\pp})_{\rm m}|^{1/2} {\cD}^{1/2}_{\rm m}  }{\cW_{\rm m}},
\ee
which yields the nonrotating limit of $\omega_I$ at $r = r_0$.

To complete our subleading analysis we also need to apply the expansion $r_{\rM} = r_0 + r_1$
to $\omega_{R}^{(1)}$. The resulting expression that follows from Eq.~(\ref{eq:omR1root1}) is
\begin{align}
    \omega_{R}^{(1)} &= \frac{\cB_{0}}{2 \omega_{\pm} \cW_{0}}
\left[ 1 + \frac{m \Omega_{0}}{2 \omega_{\pm} \cW_0}(\omega_1 - 1) \frac{\cE_{0}}{\cB_0} \right]
\nn \\
    &\quad + \left( \frac{\cB}{2 \omega_{\pm} \cW} \right)'_0 r_1.
\label{eq:omegaR_omega_indep}
\end{align}

%%%%%%%%%%%%%%%%%%%%%%%%%%%%%%%%%%%%%%%%%%%%%%%%%%%%%%%%%%%%%%%%%%%

\subsection{Subleading-order analysis with $\omega$-dependent and real $\beta_{\psi\Theta}^{(1)}$ and $\delta^{(0)}_{\psi\Theta}$}
\label{sublead_with_omega}

In this more general case, we assume expansions,
\begin{subequations}
\begin{align}
\delta^{(0)}_{\psi\Theta} & = \delta_2 (r) \ell^2 + \delta_1 (r) \ell \omega + \delta_0 (r) \omega^2,
\\
\beta_{\psi\Theta}^{(1)} & = \beta_{3}^{(1)} (r) \ell^3 + \beta_{2}^{(1)} (r) \ell^2 \omega +  \beta_{1}^{(1)} (r) \ell \omega^2,
\label{eq:delta_beta_omega_dep}
\end{align}
\end{subequations}
where the functions $\{\delta_i (r),  \beta_{i}^{(1)} (r) \}$ are taken to be real.
We can then use this prescribed $\omega$-dependence in Eq.~\eqref{subleadEq1}
and proceed exactly as described in the previous subsection to obtain
$\omega_I$ and $\omega_R^{(1)}$.
For brevity, we quote only the final results:
\begin{widetext}
\begin{align}
\omega_I &= \cF_0 \left [\,  1 - \frac{1}{V_{\rm eff}^{\pp} (r_0, \omega_0)}\left \{ 2 m \omega_0 \Omega^\p_0 (\tilde{V} \left \{ \ell^2 (1+\alpha)
+ \mu^2 r^2\right \})^\p_0 - m \omega_0 \Omega^{\pp}_0 \cW_0
+ m \omega_0 \Omega_0 (\omega_1 -1) \frac{ \cA_0}{ \cD_0}  \right \}  \, \right ]
\nn \\
&\quad + \frac{ 2 m \Omega_0\, (dr/dx)_0}{(2 \omega_0 \cW_0)^2 \cD_0^{1/2}}  |V_{\rm eff}^{\pp}(r_0,\omega_0) |^{1/2}
\left [\, \ell \omega_0 (\delta_{\rm 1})_0  \left \{ 2\omega^2_\pm (3-2\omega_1) -\tilde{V}_0 \left [ \ell^2 (1+\alpha_0) + \mu^2 r^2_0 \right ]  \right \}
+ 4 (\beta_{\psi\Theta}^{(0)})_{\rm 0} (\omega_1-1)
\right.
\nn \\
&\quad \left. -\, \ell^2 (\delta_{\rm 2})_0 (\omega_1-1) \left \{ 6\, \omega^2_0 -\tilde{V}_{\rm 0} \left [ \ell^2 (1+\alpha_0 ) + \mu^2 r^2_0 \right ] \right \}
+ \tilde{V}^2_0 (\omega_1-1) \left \{ \ell^2 (\alpha_0-1) + \mu^2 r^2_0 \right \} \left \{ \ell^2 (\alpha_0 -1) + \mu^2 r^2_0 \right \}
\right.
\nn \\
&\quad \left.
% +\, 3\omega_\pm^2 (\omega_1-1) (\lambda-1) \tilde{V}_{\rm 0} \left \{\ell^2 (\alpha_{\rm 0}-1) + \mu^2 r^2_{\rm 0} \right \}
+ (\delta_{\rm 0})_0 \left \{  2 \ell^2 \tilde{V}^2_{\rm 0} (\omega_1-3) \left ( \ell^2 \alpha_{\rm 0} + \mu^2 r^2_{\rm 0} \right )
-2 ( \beta_{\psi\Theta}^{(0)} )_{\rm 0} (\omega_1 -3)
-\, \omega^2_0 (3\,\omega_1 - 5) \tilde{V}_{\rm 0} \left [ \ell^2 (1+\alpha_{\rm 0}) + \mu^2 r^2_{\rm 0} \right ]  \right \}
\, \right ] + \cF^\p_0 \, r_1,
\nn \\
\label{eq:omegaI_omega_dep}
\\
\omega_R^{(1)} &=
\cG_0 + \frac{2 m \Omega_0}{(2 \omega_0 \cW_0)^2} (\omega_1 - 1)
\nn \\
&\quad \times \left[
-\ell \vt^3_0 (2\ell^2\alpha_0 + \mu^2 r_0^2)
\left\{ \ell^2(1+\alpha_0) + \mu^2 r_0^2 \right\}
-\ell \vt^2_0 \omega_0^2
\left\{ \ell^2 (\alpha_0^2 - 10\,\alpha_0 + 1) + (\alpha_0- 5) \mu^2 r_0^2\right\}
\right.
\nn \\
&\left.\quad +\,
\ell \vt_0 \left\{
\ell^4 (\beta_3^{(1)})_0 \, (1+\alpha_0) + \ell^3 (\beta_2^{(1)})_0\, \omega_0 (1+\alpha_0)
+ \ell^2 \left[\omega_0^2 (\beta_1^{(1)})_0\, (1+\alpha_0) + (\beta_3^{(1)})_0 \mu^2 r_0^2\right]
+ \ell (\beta_2^{(1)})_0 \mu^2 r_0^2  \right. \right.
\nn \\
&\left. \left. \quad -\, \omega_0^2
\left[ 2(1+\alpha)\omega_0^2 - \beta_1^{(1)} \mu^2 r_0^2\right]
+ \ell \omega_0 (\beta_2^{(1)})_0 \mu^2 r_0^2
\right\}
+ \frac{\ell \omega_0 \omega_1}{\omega_1 - 1}
\left\{ 2 (\beta_1^{(1)})_0 \omega_0 + \ell (\beta_2^{(1)})_0\right\} \cW_0
\right]
+ \cG_0' \, r_1.
\label{eq:omegaR_omega_dep}
\end{align}
\end{widetext}
where we defined
\begin{align}
\cG &= \frac{\ell}{2\omega_\pm \cW}
\left[
\omega_\pm^2
\left\{ \vt(1+\alpha) + \beta_1^{(1)} \right\}
- \vt^2 \left(2 \ell^2 \alpha + \mu^2 r^2 \right)
\right.
\nn \\
&\left. \quad +\, \ell^2 \beta_3^{(1)}
+ \ell \beta_2^{(1)} \omega_\pm
\right].
\end{align}
The results of the $\omega$-independent case considered in Sec.~\ref{sec:sublead_no_omega} can be recovered by setting
$\delta_0 = \delta_1 =0$ and $\beta_2^{(1)} = \beta_1^{(1)} = 0$, and making the identifications $\delta_2 = \delta^{(0)}_{\psi\Theta} \ell^{-2}$,
$\beta_{3}^{(1)} = \beta_{\psi\Theta}^{(1)}\ell^{-3}$ in Eqs.~\eqref{eq:omegaI_omega_dep}-\eqref{eq:omegaR_omega_dep}.
% %
%%%%%%%%%%%%%%%%%%%%%%%%%%%%%%%%%%%%%%%%%%%%%%%%%%%%

\section{Summary of eikonal formulae}
\label{sec:summary}

For reference, here we review the eikonal results of
Secs.~\ref{sec:leading_order}-\ref{sec:subleading_order} for the complex QNM
frequency $\omega = \omega_R + i \omega_I$ of the coupled system
\eqref{sysWaves}.

The frequency $\omega$ has the eikonal form
\be
\omega = \omega_R^{(0)} + \omega_{R}^{(1)} + i \omega_I + {\cal O}(\ell^{-1}),
\ee
where each contribution to $\omega$ has a nonrotating and rotating contributions.
For $\omega_R^{(0)}$ we have:
\be
\omega_R^{(0)} = \omega_0 + \omega_0' r_1 + m \Omega_0 \omega_1,
\ee
where $\omega_0 = \omega_{\pm}(r_0)$ is given by Eq.~\eqref{ompm} and $\omega_1$ by Eq.~\eqref{om1root1}.
As for $\omega_I$ and $\omega_{R}^{(1)}$, they are given by
Eqs.~\eqref{eq:omegaI_omega_dep} and~\eqref{eq:omegaR_omega_dep} respectively,
when the various coupling functions are allowed to have an $\omega$-dependence.
In the particular case of $\omega$-independent coupling functions, $\omega_I$
and $\omega_{R}^{(1)}$ are given by Eqs.~\eqref{eq:omegaI_omega_indep}
and~\eqref{eq:omegaR_omega_indep}, respectively.
To evaluate these expressions one needs to know the location of the effective
potential peak $r_{\rm m}$ in the nonrotating limit $r_0$ [Eq.~\eqref{eq:def_r_0}] and its
shift due to rotation $r_1$ [Eq.~\eqref{eq:def_r_1}].
The radius $r_{\rm m} = r_0 + r_1$
is defined as the common extremum of  the eikonal phase functions i.e. $(S_{,x})_{\rm m} = (H_{,x})_{\rm m} = 0$
and the corresponding potential is given by Eq.~(\ref{Ueff}).

The above results were obtained assuming the following eikonal scaling for the coupling functions,
%To obtain the above results, we assumed that the coupling functions have the eikonal scalings
%
\begin{subequations}
\begin{align}
\beta_{\psi\Theta} &= \order{4},
\qquad
\beta_{\psi} = \order{2},
\\
\gamma_{\psi\Theta} &= \order{2},
\qquad
\delta_{\psi\Theta} = \order{2},
\end{align}
\end{subequations}
where, in general, at each $\ell^{k}$-order, the coupling functions can be expanded in a sum over $\ell^{i} \omega^{j}$, such that $i + j = k$.
These scalings are a necessary condition for the proper implementation of the eikonal approximation, see discussion below
Eq.~(\ref{waveTSeik}).

To ensure consistency between two alternative methods for combining the equations of system \eqref{sysWaves}
in the eikonal limit, and that the particular limit of Paper I can be recovered,
we imposed
$(S_{,xx})_\rM = (H_{,xx})_\rM$
on the phase functions, see Eq.~\eqref{eq:ansatz}.

%%%%%%%%%%%%%%%%%%%%%%%%%%%%%%%%%%%%%%%%%%%%%%%%%%%%

\section{A more exotic class of QNM\tiny{s}}
\label{sec:exotic}

So far we have been working with the conventional eikonal scaling $\omega =
{\cal O} (\ell)$.  In this section we explore the implications of assuming a `non-Press'
QNM frequency that scales as $\omega = {\cal O} (\ell^\sigma)$ with $\sigma >1$
while keeping the same balance $\epsilon \ell = {\cal O}(1)$ between the two
eikonal parameters.  For simplicity, we also assume a \emph{nonrotating} black
hole. By suspending all previous assumptions about the eikonal order of the
coupling functions, the general expression (\ref{waveTSeik}) now yields the
leading-order equation,
\be
 \omega^4  -\beta_{\psi\Theta}^{(0)} -\frac{i}{\epsilon} \left [\,   \gamma_{\psi\Theta}^{(0)}  S_{,x} +  \delta_{\psi\Theta}^{(0)} S_{,xx} \, \right ]
+  \frac{(S_{,x})^2}{\epsilon^2}\delta_{\psi \Theta}^{(0)}  = 0,
\label{exoticEq1}
\ee
where for the reasons explained in Sec.~\ref{sec:coupledEqs} we have discarded all terms that depend
on $A,B$ and their derivatives.
%
%This expression is further reduced by evaluating it at $r_{\rm m}$,
%
Evaluating this expression at $r = r_{\rm m}$ gives
\be
 \omega^4  -(\beta_{\psi\Theta}^{(0)})_{\rm m}  -\frac{i}{\epsilon} (\delta_{\psi\Theta}^{(0)})_{\rm m} (S_{,xx})_{\rm m}  = 0.
\label{exoticEq2}
\ee
This equation offers more than one possibility for obtaining `exotic' QNMs. For instance, assuming the first two terms to
be the dominant ones [i.e.~$\beta_{\psi\Theta}^{(0)} = {\cal O} (\ell^{4\sigma})$] in the eikonal limit we have
\be
\omega^4 =  (\beta_{\psi\Theta}^{(0)})_{\rm m}.
\label{exoticEq3}
\ee
In this case we assume an expansion,
\be
\beta_{\psi\Theta}^{(0)} = \beta_{4} (r) \ell^{4 \sigma} + \beta_{3} (r) \ell^{4\sigma -1} \omega + \beta_2(r) \ell^{4\sigma -2} \omega^2.
\ee
Then,
\be
\omega^4  - \beta_{2m}  \ell^{4\sigma -2} \omega^2  - \beta_{3m}  \ell^{4\sigma -1} \omega  - \beta_{4m}  \ell^{4 \sigma} = 0.
\ee
Assuming for simplicity $\beta_3 = 0$, we find the roots
\be
\omega^2_\pm = \frac{1}{2}  \ell^{2\sigma} \left [\, \beta_{2m} \ell^{2(\sigma -1)} \pm \sqrt{ 4\beta_{4m} + \beta_{2m}^2 \ell^{4(\sigma-1)}}\, \right ].
\ee
Other possibilities include $ (\beta_{\psi\Theta}^{(0)})_{\rm m} = 0$ or $ (\delta_{\psi\Theta}^{(0)})_{\rm m}=0$, assuming each frequency-dependent
coupling term to be the dominant one in (\ref{exoticEq2}).

Clearly, the QNMs discussed in this section, assuming they exist in some theory of gravity, they should become degenerate, $\omega \to 0$,
in the GR limit of our model.
% perturbation equations.

%%%%%%%%%%%%%%%%%%%%%%%%%%%%%%%%%%%%%%%%%%%%%%

\section{Beyond spherical symmetry}
\label{sec:nonspherical}

In this section we abandon our basic assumption of spherical symmetry and go on to consider the
axisymmetric and stationary spacetime
\be
ds^2 = g_{tt} dt^2 + g_{rr} dr^2 + 2 g_{t\varphi} dtd\varphi +  g_{\theta\theta} d\theta^2 + g_{\varphi\varphi} d\varphi^2,
\ee
where $g_{\mu\nu} = g_{\mu\nu} (r,\theta)$. Given the unavailability of black hole perturbation equations for a general spacetime of
this type (even within GR) our analysis here is necessarily less ambitious than in the previous sections.

Our principal objective is to study the eikonal QNMs of the massless scalar equation,
\be
\nabla^\mu \nabla_\mu \Phi = g^{-1/2} \partial_\mu \left [ g^{1/2} g^{\mu\nu} \partial_\nu \Phi \right ] =0,
\ee
where $g = -\textrm{det}(g_{\mu\nu})$.
As we have already seen, the leading-order terms are supplied by the second-order derivatives of the  field,
\be
g^{\mu\nu}  \partial_{\mu\nu}^2 \Phi \approx 0.
\ee
Expanding this expression and assuming $\Phi \sim e^{  im\varphi-i\omega t }$ we obtain
\be
- g^{tt} \omega^2 \Phi + \frac{1}{g_{rr}} \partial_r^2 \Phi +  \frac{1}{g_{\theta\theta}} \partial_\theta^2 \Phi +
2 m\omega  g^{t\varphi} \Phi  - m^2 g^{\varphi\varphi} \Phi \approx 0,
\ee
with $g^{tt} = -g_{\varphi\varphi}/{D}$, $g^{\varphi\varphi} = -{g_{tt}}/{D}$, $g^{t\varphi} = {g_{t\varphi}}/{D}$,
% \be
% g^{tt} = -\frac{g_{\varphi\varphi}}{D}, \quad  g^{\varphi\varphi} = -\frac{g_{tt}}{D}, \quad g^{t\varphi} = \frac{g_{t\varphi}}{D},
% \ee
and $D = g_{t\varphi}^2 - g_{tt} g_{\varphi\varphi} $. Our eikonal ansatz is
\be
\Phi(r,\theta) = A(r,\theta) e^{i S(r,\theta)/\epsilon},
\ee
and at leading order we have
\be
\partial_r^2 \Phi \approx - \frac{(S_{,r})^2}{\epsilon^2} A e^{i S/\epsilon}, \quad
\partial_\theta^2 \Phi \approx - \frac{(S_{,\theta})^2}{\epsilon^2} A e^{i S/\epsilon}.
\ee
Taking $m \gg 1$ with $ \epsilon m  \sim {\cal O} (1)$ and $\omega \sim {\cal O} (m)$, the leading-order eikonal
wave equation yields
\be
g_{\varphi\varphi} \frac{\omega^2}{m^2}  - \frac{D}{\epsilon^2} \left [ \frac{(S_{,r})^2}{ g_{rr}}  +   \frac{(S_{,\theta})^2}{g_{\theta\theta}} \right ]
 + 2   g_{t\varphi} \frac{\omega}{m} + g_{tt} = 0.
 \label{leadeikaxi}
\ee
Assuming the existence of a point $(r,\theta)=(r_0,\theta_0)$  where $S_{,r} = S_{,\theta}=0$, this equation becomes a binomial
for $\omega/m$:
\be
Q (r_0,\theta_0,\omega)  \equiv \left (\, g_{\varphi\varphi}  \frac{\omega^2}{m^2} + 2 g_{t\varphi}  \frac{\omega}{m}   + g_{tt} \, \right )_0 = 0,
\label{Qeq1}
\ee
with solutions
\be
\omega_\pm = m \left (\frac{ - g_{t\varphi} \pm D^{1/2}}{g_{\varphi\varphi}} \right )_0.
\label{qnmrot}
\ee
Following the logic of the spherical calculation, the equations for $(r_0,\theta_0)$ can be derived by taking the
$r$ and $\theta$ derivatives of  (\ref{leadeikaxi}) and setting $\omega=\omega_\pm$, i.e.
\be
Q_{,r} (r_0,\theta_0,\omega_\pm ) = 0 =  Q_{,\theta} (r_0,\theta_0,\omega_\pm ).
\label{Qeq2}
\ee
Inspection of these equations reveals that the QNMs (\ref{qnmrot}) are the ones associated with the spacetime's photon rings, equatorial
or nonequatorial alike, see \cite{Glampedakis_Pappas_2019} for more details. Indeed, the geodesic potential for photons is
\be
V_{\rm geod} (r,\theta, b)= \frac{1}{D} \left (\, g_{tt} b^2 + 2 g_{t\varphi} b + g_{\varphi\varphi} \, \right ),
\ee
where $b$ is the impact parameter. With the familiar eikonal identification
\be
\omega_\pm = \frac{m}{b},
\ee
the preceding conditions (\ref{Qeq1}) and (\ref{Qeq2}) are equivalent to
\be
V_{\rm geod}=0, \quad V_{{\rm geod},r} =0, \quad V_{{\rm geod},\theta} =0,
\ee
that describe a photon ring of radius $r=r_0$ located at a latitude  $\theta=\theta_0$. For example, as discussed in
Ref.~\cite{Glampedakis_Pappas_2019}, such photon rings appear above some spin threshold in two very familiar nonseparable
stationary-axisymmetric metrics, the Johannsen-Psaltis~\cite{Johannsen:2011dh} and Hartle-Thorne spacetimes~\cite{Hartle1967,Hartle:1968si}.
In these two examples the emergence of nonequatorial photon rings appears to be closely related
to the two-lobed shape of the event horizon~\cite{Glampedakis_Pappas_2019}.

Equation~(\ref{leadeikaxi}) could admit more solutions than the ones associated with photon rings. An example is provided
by the Kerr spacetime where the `nonequatorial' QNMs with $|m| < \ell$ are known to be associated with nonequatorial
spherical photon orbits in the eikonal limit \cite{Dolan:2010wr, Yang:2012he}. Given the intimate relation between the existence
of spherical orbits and the separability of the geodesic equations (see Ref.~\cite{Glampedakis_Pappas_2019} for a detailed discussion)
it is not surprising that the separability of (\ref{leadeikaxi}) is the key reason behind the presence of another family of eikonal QNM solutions.

As a case in point we hereafter assume a Kerr metric and write $S(r,\theta) = S_r (r) + S_\theta (\theta)$.
Then Eq.~(\ref{leadeikaxi}) indeed separates, leading to the pair of equations,
\begin{align}
(\Delta  S_{r,r})^2 &= a^2 + \Delta C  + \frac{1}{2} \lambda^2 \left [\, 2 r^4 + a^4 +  (3r + 2M) a^2 r \, \right ]
\label{radial1}
\nn \\
&\quad  - 4aM r \lambda \equiv R(r,\lambda, C),
\\
(S_{\theta,\theta})^2 &= - \left ( C + \frac{1}{\sin^2\theta} \right ) + a^2 \lambda^2 \left ( \cos^2\theta -\frac{1}{2} \right ),
\label{angular1}
\end{align}
where $C$ is the separation constant and $\lambda \equiv \omega/m$.  It is easy to see that the angular equation can be matched
term by term to the angular photon geodesic equation,
\be
u_\theta^2 = \cQ - b^2 \cot^2 \theta + a^2 \cos^2 \theta,
\ee
where $\cQ = Q/E^2$ is the rescaled Carter constant and $b = L/E$ is the impact parameter ($E,L$ are the orbital energy and angular momentum
per unit mass). The two equations are identical if we identify
\be
\lambda = \frac{1}{b}, \quad C = -1 - \lambda^2 \left  ( \cQ + \frac{1}{2} a^2 \right ), \quad
\frac{S_{\theta,\theta}}{\lambda} = u_\theta.
\ee
Using the same relations between eikonal and geodesic parameters we can show that the radial equation (\ref{radial1})
exactly matches the radial geodesic photon equation~\cite{Bardeen:1972fi}.

Assuming a radius $r=r_0$ where $S_{r,r} (r_0)=0$, we have
\begin{align}
R_0 &=  \lambda^2 \left [\, r^4_0 + (r_0+2M) a^2 r_0  - \Delta_0 \cQ \,\right ] - 4 aM r_0 \lambda
\nn \\
&\quad  - r_0 (r_0-2M)= 0,
\nn\\
\frac{1}{2} R^\p_0 &=    \lambda^2 \left [\, 2 r^3_0 +  (r_0 + 2M) a^2   - (r_0 -M) \cQ \,\right ]
\nn \\
&\quad - 2 aM  \lambda + M - r_0  =0.
\end{align}
These are the same equations that describe a spherical photon orbit of radius $r_0$ with parameters $b,\cQ$
given by the previous relations. The best way of proceeding is to work with the combination $R_0 - r_0 R^\p_0 = 0$.
This yields
\be
\lambda^2_0 =   \frac{r_0^2}{ r_0^2 (3 r_0^2 + a^2 ) - (r_0^2 -a^2 ) \cQ}.
\label{lam0}
\ee
We can subsequently  solve $R_0=0$ with respect to the linear $\lambda$ term, take its square and substitute
$\lambda^2 \to \lambda^2_0  $. The outcome of this exercise is
\be
\cQ_0 = \frac{r^3_0 [ \, 4 a^2 M - r_0 (r_0 -3M)^2 \, ]}{a^2 (r_0 -M)^2}.
\ee
Using this in (\ref{lam0}),
\be
\lambda_0  = \frac{\omega_0}{m} = \frac{a (r_0 -M)}{r_0^2 (r_0 -3M) + a^2 (r_0 + M)}.
\ee
These expressions, with their exact correspondence to spherical photon orbits, represent the leading-order eikonal formulae for nonequatorial
Kerr QNMs. An analysis along these lines could be carried out for any other separable spacetime,
for example, the deformed Kerr metric of Ref.~\cite{Johannsen2013PhRvD}.

%%%%%%%%%%%%%%%%%%%%%%%%%%%%%%%

\section{Concluding remarks}
\label{sec:conclusions}

This paper's eikonal analysis of the fundamental QNM of spherically symmetric black holes generalises the results of
Paper I~\cite{Glampedakis:2019dqh} by considering additional couplings between the tensorial gravitational field and the
(potentially massive) scalar field and, to some extent, the presence of slow rotation.

The resulting eikonal expressions, although somewhat unwieldy and markedly more complicated than those of Paper I, should encompass a large
class of black holes beyond GR. The algebraic complexity of our results is the main reason we have not attempted a similar analysis (although
such analysis should be feasible) for the coupled system of three black hole perturbation equations that appear in generalised
vector-tensor~\cite{Tattersall_etal2018} and in Einstein-Maxwell-dilaton theories~\cite{Brito:2018hjh}.
As was the case for the perturbation equations of Paper I, we have been able to show
that the eikonal QNM can be associated with the peak of an effective potential although no correspondence to a geodesic photon ring
appears to exist (this issue was explored in detail in~\cite{Glampedakis:2019dqh}).

As a secondary objective of our eikonal analysis, we have explored the possibility of having `exotic' QNMs; the frequency of such
modes is only a function of the coupling parameters that appear in the perturbation equations and becomes trivial (i.e. $\omega \to 0$)
in the GR limit.

Our toy model of eikonal scalar QNMs in a stationary-axisymmetric black hole spacetime has revealed a number of interesting
properties. There is a class of modes associated with geodesic photon rings; these rings are typically found in the equatorial
plane (as in the Kerr spacetime) but could also appear in pairs off the equatorial plane (as in, for example, the Johannsen-Psaltis
spacetime~\cite{Johannsen:2011dh}). As thoroughly discussed in Ref.~\cite{Glampedakis_Pappas_2019}, these nonequatorial photon rings
are likely to be the hallmark of geodesically nonseparable spacetimes describing non-Kerr black holes. As far as we are aware, this paper
provides the first calculation (albeit within the eikonal approximation) of the QNMs associated with these photon rings. If black holes with
nonequatorial photon rings exist, their ringdown signals could show interference between the ringdown signals of the
individual photon rings.

The second class of QNMs in stationary-axisymmetric black holes is the one associated with nonequatorial spherical photon orbits.
These orbits/modes appear provided the spacetime is separable with a Carter-like constant; in the case of Kerr black holes,
they correspond to the well-known `nonequatorial' $|m| < \ell$ modes. Similar QNMs could characterize the black holes of dynamical Chern-Simons gravity
where geodesic motion has been conjectured to be endowed with a higher-than-two rank Killing tensor~\cite{Cardenas-Avendano:2018ocb}.

The present work could be extended in a number of ways. The most obvious one is the application of our eikonal formulae to
black holes of particular theories of gravity where QNM numerical data are available. Unfortunately, a full numerical solution of
the wave equations considered in this paper is still lacking in the literature. The only comparison performed so far concerns
the less general equations of Paper I which can be used to model the QNMs of Schwarzschild black holes in dynamical Chern-Simons gravity.
The outcome of this exercise has demonstrated the relatively high accuracy of the eikonal results.
Another interesting possibility is to implement the eikonal formulae as a generic parametrised scheme with the aim of constraining the coupling between the
tensor and scalar degrees of freedom. Thinking further ahead, the eikonal techniques of this paper should be directly applicable to the
perturbation equations describing rapidly rotating black holes beyond GR when such equations are derived at some point in the future.

%%%%%%%%%%%%%%%%%%%%%%%%%%%%%%%%%%%%%%%%%%%%%%%%%%%%

\acknowledgements

H.O.S~acknowledges financial support through NASA Grants No.~NNX16AB98G and
No.~80NSSC17M0041.  K.G.~acknowledges financial support from Grants
No.~FIS2015-3454 by the Spanish Ministerio de Educaci\'on y Ciencia and
No.~PI2019-2356B by the Fundaci\'on Seneca (CARM Murcia) and networking support by
the COST Actions GWverse CA16104 and PHAROS CA16214.

%%%%%%%%%%%%%%%%%%%%%%%%%%%%%%%%%%%%%%%%%%%%%%%%%%%%

%\appendix

%%%%%%%%%%%%%%%%%%%%%%%%%%%%%%%%%%%%%%%%%%%

\bibliography{biblio.bib}

\end{document}